# 2D Barcode for DNA Encoding

### Elena Purcaru, Cristian Toma

*Bucharest General Medicine Faculty, Cybernetics and Economic Informatics Faculty*
*"Carol Davila" University of Medicine and Pharmacy, Academy of Economic Studies*
*Eroii Sanitari Boulevard 8, Bucharest, Romana Square 6, Bucharest*
*ROMANIA*
*elena.purcaru@gmail.com, cristian.toma@ie.ase.ro*

**Abstract:** The paper presents a solution for endcoding/decoding DNA information in 2D barcodes. First part focuses on the existing techniques and symbologies in 2D barcodes field. The 2D barcode PDF417 is presented as starting point. The adaptations and optimizations on PDF417 and on DataMatrix lead to the solution − DNA2DBC − DeoxyriboNucleic Acid Two Dimensional Barcode. The second part shows the DNA2DBC encoding/decoding process step by step. In conclusions are enumerated the most important features of 2D barcode implementation for DNA.

## 1. Introduction

A barcode [6] is an optical representation of data. Originally, barcodes represented data as parallel lines and the spacings, referred to as linear or 1D (1 dimensional) barcodes or symbologies. They also come in patterns of squares, dots, hexagons and other geometric patterns within images termed 2D (2 dimensional) matrix codes or symbologies. Although 2D systems use symbols other than bars, they are generally referred to as barcodes as well. [6]

Barcodes can be read by optical scanners called barcode readers, or scanned from an image by special software. A barcode reader contains a photo-sensor that converts the barcode into an electrical signal as it moves across it. The scanner then measures the relative widths of the bars and spaces, translates the different patterns back into regular characters, and sends them to a computer or portable terminal. 2D readers are based mostly on camera with CMOS sensor picture processing technology.

Barcodes were invented to label railroad cars, but they were not commercially successful until they were used to automate supermarket checkout systems, a task in which they have become almost universal. [6].

## 2. Barcodes Types

Each character in a barcode is represented by a pattern of wide and narrow bars. Every barcode begins with a special start character and ends with a special stop character [5-6]. These conventions help the barcode scanner to identify and read the symbol in the right position. Some barcodes may include a checksum character. A checksum is calculated when the barcode is printed using the characters in the barcode. The reader performs the same calculation in order to detect errors in the symbol. If the two checksums don't match, the reader assumes that something is wrong, throws out the data, and tries again.

### 2.1. 1D Numeric and Alphanumeric Barcodes

A Barcode Symbology [5-6] defines the technical details of a particular type of barcode: the width of the bars, character set, method of encoding, checksum specifications, etc. Barcode users are usually interested in the general capabilities of a particular symbology (how much and what kind of data can it hold, what are its common uses, etc) and not in the technical details.

Most used 1D numeric barcodes /symbologies are:

- Codabar: used in library systems, sometimes in blood banks
- Code 11: used primarily for labeling telecommunications equipment
- EAN-13: European Article Numbering international retail product code
- EAN-8: compressed version of EAN code for use on small products
- Industrial 2 of 5: older code not in common use anymore
- Interleaved 2 of 5: widely used in industry, air cargo
- Plessey: older code commonly used for retail shelf marking
- MSI: variation of the Plessey code commonly used in USA
- PostNet: used by U.S. Postal Service for automated mail sorting
- UPC-A: Universal Product Code seen on almost all retail products in the USA and Canada
- Standard 2 of 5: older code not in common use
- UPC-E: compressed version of UPC code for use on small products.

Most used 1D alphanumeric barcodes /symbologies are:

- Code 128: very capable code, excellent density, high reliability; in very wide use world-wide
- Code 39: general-purpose code, wide use world-wide
- Code 93: compact code similar to Code 39
- LOGMARS: same as Code 39, is the U.S. Government specification

## 2.2.  2D Barcodes

Two dimensional – 2D symbols encode data in two dimensional shapes.  They fall into two general categories:

- Stacked barcodes, constructed like a layer of barcodes stacked on top of the other; they can be read by special 2D scanners or by many CCD (charge-coupled device) and laser scanners with the aid of special decoding software.
- Matrix Codes, built on a true 2D matrix; they are usually more compact than a stacked barcode, and

they can be read only by 2-D scanners.

The main advantage of 2D barcodes is the ability to encode a lot of information in a small space.  If 1D barcodes can encode 20 to 25 characters, 2D symbols can encode from 100 to about 2,000 characters.

The most used 2D barcodes / symbologies are (Classification according to [5]):

- PDF417: used for encoding large amounts of data
- DataMatrix: can hold large amounts of data, in very small codes
- Maxicode:  fixed length, used by United Parcel Service for automated package sorting
- QR Code:  Used for material control and order confirmation
- Aztec
- Data Code
- Code 49

## 3. Barcode 2D – PDF417 Analysis

The PDF 417 code is part of 2 dimentional barcode family. PDF stands for "Portable Document File" because with several rows and columns, it is possible to encode up to 2700 bytes - *a lot of PDF417 info is copyrighted in [4]*.

The encoding is done in two stages:

- ***High level encoding*** – The datas (input bytes) are converted to "codeword". From now on CW stands for codeword. ***High level encoding*** supports multiple modes encoding such as:
  - *Byte* level – capacity of encoding is ASCII code 0 to 255, aprox. 1,2 byte per CW. The "Byte" mode (high level encoding) allows encoding 256 different bytes, which is the entire extended ASCII table (ISO 8859). For ASCII code values please reffer [8].
  - *Text* level – capacity of encoding is ASCII code 9, 10, 13 and 32 to 127,  2 characters per CW

- *Numeric* level – capacity of encoding is only for digits 0 to 9, 2.9 digits per CW
- ***Low level encoding*** – The codewords obtained during first stage are converted to bars and spaces patterns.

Moreover an error correction system with several levels is included in order to allow reconstituting badly printed, erased, fuzzy or torn off datas.

The general structure of PDF 417 is [4] – *CW stands for codeword*:
- The width of the smalest/finest bar is called the module.
- A bar module is represented by "1" and a space module by "0".
- The code has 3 to 90 rows.
- A row has 1 to 30 datas columns and its width goes from 90 to 583 modules with the margins.
- Maximum number of CW in bar codes: 928 including 925 for the datas. (1 for the length descriptor and 2 at least for the error correction.)

- There are 929 CWs including 900 for the datas, they are numbered from 0 to 928.
- The errors correction levels goes from 0 to 8. The correction covers 2 (on level 0) to 512 (on level 8) CW.
- The row consists of: "a start character", "a left side CW", "1 to 30 datas CW", "a right side CW" and "a stop character". There must be a white margin of at least 2 modules on each side.
- CW of padding (e.g. codeword with value "900") can be intercalated between datas and correction CW; those must be located at the end.
- First CW indicates CW total number of the code including: datas, CW of stuffing and itself but excluding CW correction.
- "Macro PDF417" mechanism allows distributing more datas on several bar codes.

The CW number 900-928 have special meaning, some enable to switch between modes in order to optimise the code-Table1.

*Table 1.Special CW*

| CW number | : Function |
|---|---|
| | |
| 900 | : Switch to "Text" mode |
| 901 | : Switch to "Byte" mode |
| 902 | : Switch to "Numeric" mode |
| 903 to 912 | : Reserved |
| 913 | : Switch to "Octet" only for the next CW |
| 914 to 920 | : Reserved |
| 921 | : Initialization |
| 922 | : Terminator codeword for Macro PDF control block |
| 923 | : Sequence tag to identify the beginning of optional fields in the Macro PDF control block |
| 924 | : Switch to "Byte" mode (If the total number of byte is multiple of 6) |
| 925 | : Identifier for a user defined Extended Channel Interpretation (ECI) |
| 926 | : Identifier for a general purpose ECI format |
| 927 | : Identifier for an ECI of a character set or code page |
| 928 | : Macro marker CW to indicate the beginning of a Macro PDF Control Block |

In this section is presented the practical encoding of word "Super!" in high-level both in text and byte mode. This presentation is support for our 2D barcode defined for encoding DNA structure.

The ***high-level encoding in text mode*** has 4 sub-modes:
- Uppercase
- Lowercase
- Mixed: Numeric and punctuation
- Punctuation

The default sub-mode is "Uppercase", in this sub-mode 2 characters are encoded in each CW, here is the characters table 2:

*Table 2.Characters value for high-level encoding in text mode*

| Value | Uppercase | Lowercase | Mixed | Punctuation |
|---|---|---|---|---|
| 0 | A | a | 0 | ; |
| 1 | B | b | 1 | < |
| 2 | C | c | 2 | > |
| 3 | D | d | 3 | @ |
| 4 | E | e | 4 | [ |
| 5 | F | f | 5 | \ |
| 6 | G | g | 6 | ] |
| 7 | H | h | 7 | _ |
| 8 | I | I | 8 | ` (Quote) |
| 9 | J | j | 9 | ~ |
| 10 | K | k | & | ! |
| 11 | L | l | CR | CR |
| 12 | M | m | HT | HT |
| 13 | N | n | , | , |
| 14 | O | o | : | : |
| 15 | P | p | # | LF |
| 16 | Q | q | - | - |
| 17 | R | r | . | . |
| 18 | S | s | $ | $ |
| 19 | T | t | / | / |
| 20 | U | u | + | " |
| 21 | V | v | % | | |
| 22 | W | w | * | * |
| 23 | X | x | = | ( |
| 24 | Y | y | ^ | ) |
| 25 | Z | z | PUN | ? |
| 26 | SP | SP | SP | { |
| 27 | LOW | T_UPP | LOW | } |
| 28 | MIX | MIX | UPP | ' (Apostrophe) |
| 29 | T_PUN | T_PUN | T_PUN | UPP |

The 6 switchs are included in these tables, they allow to change the sub-mode:

- UPP: switch to "Uppercase"
- LOW: switch to "Lowercase"
- MIX: switch to "Mixed"
- PUN: switch to "Punctuation"
- T_UPP: switch to "Uppercase" only for next character
- T_PUN: switch to "Punctuation" only for next character

Each CW encodes 2 characters; if $C_1$ and $C_2$ are the values of the two characters, CW value is: $C_1 * 30 + C_2$

If it remains an alone character, we add to it a padding switch, for instance T_PUN.

For sample encoding of word "Super" in high-level text mode please see Tabel 3.

*Table 3. Text Mode High-level Encoding*

Sample, word to encode: "Super !"

S : 18, LOW : 27, u : 20, p : 15, e : 4, r : 17, SPACE : 26, T_PUN : 29, ! : 10
that is 9 characters, it will be a T_PUN for the padding.
CW0 = 18 * 30 + 27 = 567
CW1 = 20 * 30 + 15 = 615
CW2 =  4 * 30 + 17 = 137
CW3 = 26 * 30 + 29 = 809
CW4 = 10 * 30 + 29 = 329
The sequence is consequently : 567, 615, 137, 809, 329

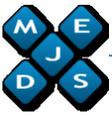

**High-level encoding in byte mode** structure is:

- If the byte number is a multiple of 6, it will be used the 924 CW to switch to "Byte" mode; if the byte number is not a multiple of 9, it will be used 901 CW for switching.

- The coding consists has to transform 6 bytes in base 256 to 5 CW in base 900. The conversion has several steps:

  - Take the bytes by group of 6; let $X_5$ to $X_0$ their decimal values. ($X_0$ is the less significant character)

  - Compute the sum $S = X_5 * 256^5 + X_4 * 256^4 + X_3 * 256^3 + X_2 * 256^2 + X_1 * 256 + X_0$

  - Compute the CWs: CW0 = S MOD 900, new value of S : S = S DIV 900, CW1 = S MOD 900 and so forth to CW4 (CW0 is the less significant CW)

  - Bytes which remain after the conversion of the groups of 6 are taken just as they are: 1 byte = 1 CW of same value.

In table 4 there is the word "Super!" encoded in byte mode:

*Table 4. Word "Super!" high-level encoded in byte mode*

| |
|---|
| **Word:** *Super!*<br><br>The sequence of bytes (in ASCII) is: 83, 117, 112, 101, 114, 33<br><br>$S = 83 * 256^5 + 117 * 256^4 + 112 * 256^3 + 101 * 256^2 + 114 * 256 + 33 =$<br>$S = 91259465105408 + 502511173632 + 1879048192 + 6619136 + 29184 + 33 = 91763861975585 = 917\ 638\ 619\ 755\ 85$<br><br>CW0 = 917 638 619 755 85 MOD 900 = 485<br>S = 917 638 619 755 85 DIV 900 = 101959846639 .5388888888888888889<br><br>CW1 = 101959846639 MOD 900 = 439<br>S = 101959846639 DIV 900 = 113288718 .4877777777777777777778<br><br>CW2 = 113288718 MOD 900 = 318<br>S = 113288718 DIV 900 = 125876 .35333333333333333333333333<br><br>CW3 = 125876 DIV 900 = 776<br>S = 125876 DIV 900 = 139 .862222222222222222222222222 |

CW4 = 139 MOD 900 = 139

The sequence including the switch (CW with value 924) is consequently for word "Super!":
924, 139, 776, 318, 439, 485

No matter what kind of high-level engoding has been done (text/byte/numeric mode), Left Side CW and Right Side CW are computed according to the table used for the actual row.

To obtain the CW value, make the following calculation (where '/' is DIV – integers division): (Row Number / 3) * 30 + X with X taken in the following table: (Firts row is row number 0)

*Table 5. Tables formula used to encode*

| Table used to encode the CWs of this row | For the left side CW | For the right side CW |
|---|---|---|
| 1 | (Rows No -1) / 3 | Data columns No - 1 |
| 2 | (Security level * 3) + (Rows No- 1) MOD 3 | (Rows No -1) / 3 |
| 3 | Data columns No - 1 | (Security level * 3) + (Rows No - 1) MOD 3 |

**Low-level encoding** general structure is (no matter if in high-level encoding is done in byte/text/numeric mode):

- Each CW is made of 17 modules, containing 4 bars and 4 spaces (Name code comes from there!) and it start by a bar. Bars and spaces width is 1 to 6 modules. (Except for start and stop characters) Sample for CW = 11101010001110000:

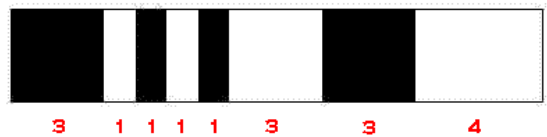

- Start character is: 11111111 0 1 0 1 0 1 000

- Stop character is: 1111111 0 1 000 1 0 1 00 1 (Here there is a 5th bar, thus 18 modules.)
- There are 3 distinct tables for encoding the 929 codewords.
- Each row uses only one encoding table, this table will be used again 3 rows further. Sample: Row 1 -> table 1, row 2 -> table 2, row 3 -> table 3, row 4 -> table 1, ... , etc.

There is the formula: table number = ( ( "row number" MOD 3 ) * 3 )

The 3 tables giving the patterns for the 929 codewords start like this (all 929 codewords are distributed in unique manner into 3 tables, e.g. CW = 11010101111100000 can be found only in $3^{rd}$ table first position – Table 6):

Table 6 – The tables for CWs

| Table 1 | Table 2 | Table 3 |
|---|---|---|
| 111 0 1 0 1 0 111 000000 | 11111 0 1 0 1 0 11 00000 | 11 0 1 0 1 0 11111 00000 |
| 1111 0 1 0 1 0 1111 0000 | 111111 0 1 0 1 0 111 000 | 111 0 1 0 1 0 111111 000 |
| 11111 0 1 0 1 0 11111 00 | 1111 0 1 0 1 00 1 000000 | 1 0 1 0 1 00 1111 000000 |
| 111 0 1 0 1 00 111 00000 | 11111 0 1 0 1 00 11 0000 | 11 0 1 0 1 00 11111 0000 |
| ..... | ..... | ..... |

Figure 1 presents the word "Super!" in PDF 417 encoded at high-level in byte mode with 2 columns and security level 1:

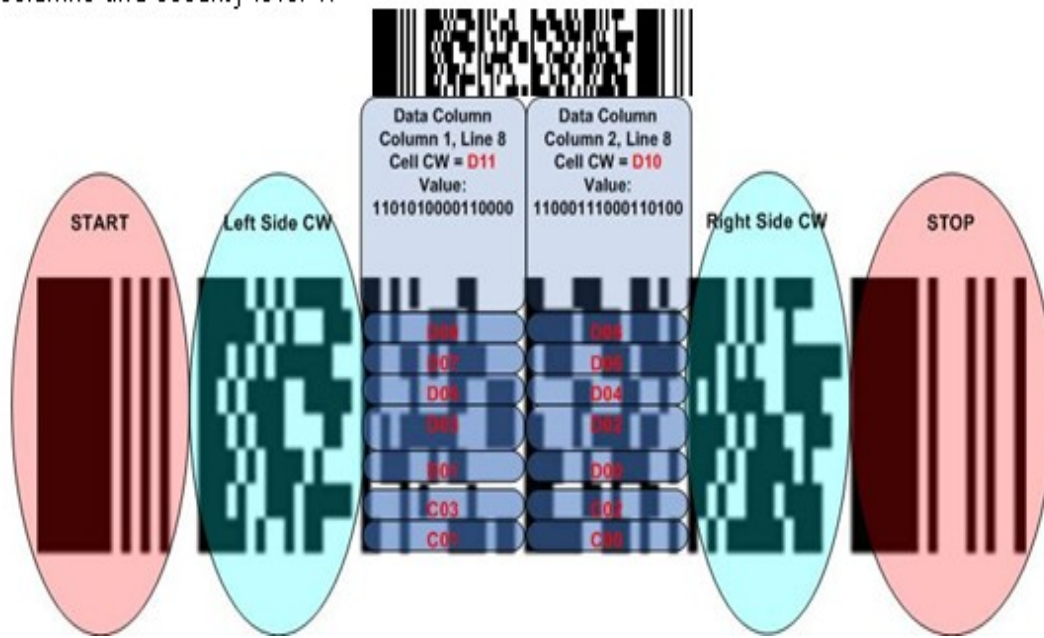

*Figure 1. PDF 417 for word "Super!" at byte level within 2 columns.*

In table 7 there is the encoding of word "Super!" in 2 columns with D0…D10 codewords and correction code C0…C4.

Table 7 – Encoding/decoding low-level

| The sequence including the switch (CW with value 924) is consequently for word "Super!": |
|---|
| **924, 139, 776, 318, 439, 485** |

| | |
|---|---|
| 11010100001100000-Table 1, Line 008+1 | 11000111000110100-Table 1, Line **924**+1 |
| 11010001110100000-Table 2, Line **139**+1 | 11111011110111010-Table 2, Line **776**+1 |
| 11001101011110000-Table 3, Line **318**+1 | 10111000111011110-Table 3, Line **439**+1 |
| 11001000011001110-Table 1, Line **485**+1 | 10000110001100100-Table 1, Line 900+1 |
| 10101111110001110-Table 2, Line 065+1 | 11000101111011000-Table 2, Line 482+1 |
| 10111001011000000-Table 3, Line 393+1 | 11110010011110010-Table 3, Line 214+1 |
| 10001010000100000-Table 1, Line 364+1 | 11100011010011000-Table 1, Line 620+1 |
| 11001000011110110-Table 2, Line 420+1 | 11111100110101010-Table 2, Line 729+1 |
| D11=008, | D10=**924** |
| D09=**139**, | D08=**776** |
| D07=**318**, | D06=**439** |
| D05=**485**, | D04=out of scope |
| D03=out of scope, | D02=out of scope |
| D01=out of scope, | D00=out of scope |
| C03=364, | C02=620 |
| C01=420, | C00=729 |

CW with value D11 (11010100001100000) is a code which can be found in Table 1 at line 9 – because lines starts from 0 in the tables the row 9 is actually 8. This means the CW D11 has corresponding value 8. The D11 codeword represents the length in codewords of the low-level encoding scheme. The CW D10 with value 11000111000110100 which is correspondence value 924 (row 925[th] from table 1) means "switching to byte level encoding". After this D05, D06, D07, D08 and D09 encodes the codewords value by value.

*Table 8. Correction System*

| Level | Number of CWs required by the correction system, 2 of which for the detection ( $2^{level + 1}$ ) | Maximum number of data CWs |
|---|---|---|
| 0 | 2 | 925 |
| 1 | 4 | 923 |
| 2 | 8 | 919 |
| 3 | 16 | 911 |
| 4 | 32 | 895 |
| 5 | 64 | 863 |
| 6 | 128 | 799 |
| 7 | 256 | 671 |
| 8 | 512 | 415 |

**The correction codes are computed based on Reed-Solomon codes**. The general error correction and detection procedure is:

- Error detection system use 2 CWs and correction system use some between 2 and 510.

- The number of CWs to add depend of the correction level used, because of the limit to 928 CWs in a bar code (1 of which for the sum of CWs) the maximum level is limited by the number of data CWs. The number of CWs that the error correction algorithm can reconstitute is equal to the number of CWs required by the correction system.

- The recommended correction level depends on the number of data CWs:

*Table 9. Recommended correction level*

| Number of data CWs | Recommended correction level |
|---|---|
| 1 up to 40 | 2 |
| 41 up to 160 | 3 |
| 161 up to 320 | 4 |
| 321 up to 863 | 5 |

- Reed Solomon codes are based on a polynomial equation where x power is $2^{s+1}$ with s = error correction level used. For instance with the correction level 1 it should be used an equation like this: $a + b*x + c*x^2 + d*x^3 + x^4$. The numbers a, b, c and d are the factors of the polynomial equation. The factors look like:

*Table 10. Factors of polynomial equation*

| Factor (coefficients) table for level 0 |
|---|
| 27 917 |
| Factor (coefficients) table for level 1 |
| 522 568 723 809 |
| Factor (coefficients) table for level 2 |
| 237 308 436 284 646 653 428 379 |
| Factor (coefficients) table for level 3 |
| 274 562 232 755 599 524 801 132 295 116 442 428 295  42 176 65 |
| Factor (coefficients) table for level 4 |
| 361 575 922 525 176 586 640 321 536 742 677 742 687 284 193 517 |
| 273 494 263 147 593 800 571 320 803 133 231 390 685 330  63 410 |
| ... |

- Let '**s**' the correction level used, '**k**' = $2^{s+1}$ the number of correction CWs, '**a**' is the factors/coefficients array, '**m**' the number of data CWs, '**d**' the data CWs array and '**c**' the correction CWs array. There is a temporary variable '**t**'. '**c**' and '**t**' are inited with 0. The C/C++/Java source code is:

*Table 11. Computing Correction CWs array*

```
for(int i = 0; < m; i++) {
    t = (d[i] + c[k - 1]) % 929;
    for(int j = k - 1; j == 0; j--)
{

    if( j == 0 ) {
        c[j] = (929 - (t * a[j]) %
929) % 929;
    } else {
    c[j] = (c[j - 1] + 929 - (t *
a[j]) % 929) % 929;
    }
    }
}

for(int j = 0; j < k; j++)
    if( c[j] != 0)
        c[j] = 929 - c[j];
```

After this, the one can apply the "recipe" by using the obtained codes in the reverse order (From last to first).

# 4. DNA2DBC – DeoxyriboNucleic Acid Two Dimensional Barcode

In this section we present a solution for coding DNA into 2D barcode.

## 4.1. DNA Coding

Any genetic information (species, hair colour, number of limbs etc) is encoded as DNA (Deoxyribonucleic acid). DNA is represented through symbols – nitrogenous bases or nucleotides. All these symbols form the genetic alphabet:  A, T, G, C. Three associated symbols form code words: the codons. The sequence of several code words describes logical sentences: proteins (peptides and polypeptides). These sentences are combined to convey a message – a trait. DNA therefore contains all information in a living organism. RNA (Ribonucleic Acid) is very similar to DNA. DNA's nitrogenous base timine(T) is replaced with uracil (U).

## 4.2. Barcode DNA

DNA information is nowadays obtained through sequencing process. That determines DNA from a tissue sample (blood, hair etc). The result of a sequencing operation is a given in text format, and in graphical form, but DNA sequences is mainly stored as text files, for portability. A DNA Sequencer can read more than 2 million nucleotides in 24 hours. Although there can be read fragments of up to 1.000 nucleotides, modern sequencers are able to operate with several sequences simultaneously, which significantly raises their capacity and productivity. In 2003 Paul Hebert suggested using bar coding techniques for organization of species. The barcode assigned is based on the CO1 gene.[6] The initiative was contested by various religious groups. However, the idea of representing DNA with barcodes was born and nowadays it finds justification in the latter 2D barcode representation. Being able to encode a lot of information in a small space, 2D barcodes are a great solution in bringing the hundreds/thousands DNA symbols closer to their users. It will be enough to read a 5 cm square shape from a piece of paper to have DNA data otherwise occupying half of page or registered in a Database among other million records. These shapes instead, can be placed on medical files, laboratory recipients, taking little space a giving large info in the shortest time.

## 4.3. Barcode Encoding Scheme

This section presents a solution for DNA encoding/decoding as two dimensional barcode - 2D matrix. The DNA string will be translated into 3 bit code words (CW), according to proprietary simbology. The result code will then be added a checksum for error detection. Correction CWs will be further added, making possible to recover lost DNA data. Data can be lost in the printing process or by physical damaging of the printed code (scratching, staining etc.), that makes the code symbols not fully readable. Further on, the 3bits CWs will be split into rows, matrix positioning patterns will be added and the resulting matrix will be converted into one binary string. This string is printable as 2D barcode with individual drawing of each bar using a graphical interface, or with custom fonts using a text editor.

The decoding process is the reverse of the one described above: a graphical symbol will be translated into binary string based on bars and spaces, and the process continues, the string developing into a DNA sequence.

The following sections will refer only to the encoding process, since decoding means applying the algorithm in reverse.

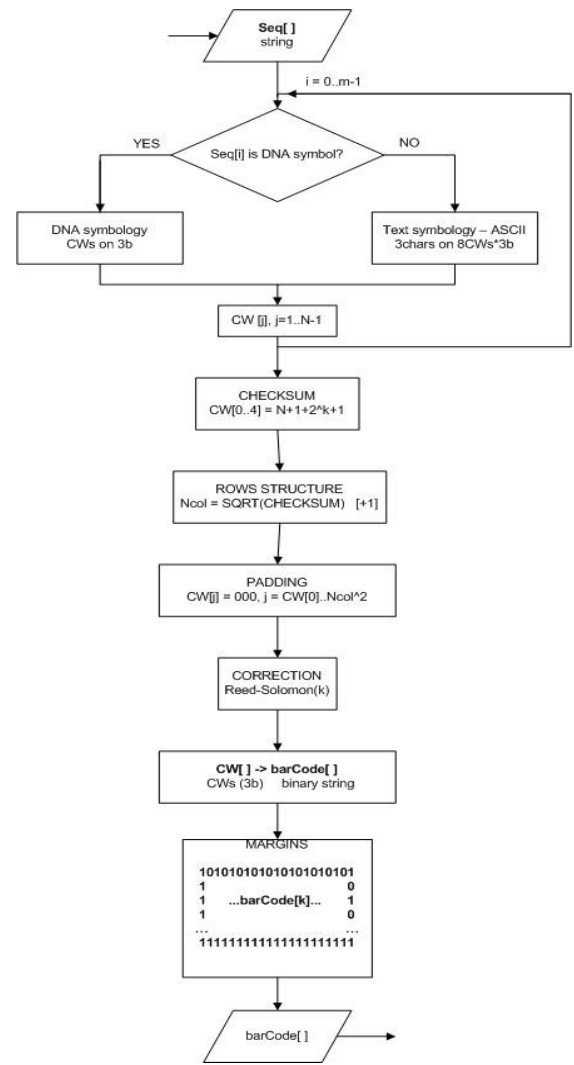

*Figure 2. DNA to 2D barcode logical chart flow*

## 4.4. Input and Output

The algorithm will have as input a DNA string. The main purpose is to translate this sequence into a string printable as bar code symbols.

A DNA string is basically a string of only four possible symbols (A, C, G, T/U). But in practice, DNA sequence rarely comes alone, but having some explaining data (metadata) about sequence origin, method used to identify it etc. Therefore the DNA input comes in two types of data: text and real DNA, as observed in a sample of DNA FASTA file in table 12:

*Table 12. DNA FASTA file*

```
>AB000263
|acc=AB000263|descr=Homo sapiens
mRNA for prepro cortistatin like
peptide, complete cds.|len=368
ACAAGATGCCATTGTCCCCCGGCCTCC
TGCTGCTGCTGCTCTCCGGGGCCACGG
CCACCGCTGCCCTGCCCCTGGAGGGTG
GCCCCACCGGCCGAGACAGCGAGCATA
TGCAGGAAGCGGCAGGAATAAGGAAAA
GCAGCCTCCTGACTTTCCTCGCTTGGT
GGTTTGAGTGGACCTCCCAGGCCAGTG
CCGGGCCCCTCATAGGAGAGGAAGCTC
GGGAGGTGGCCAGGCGGCAGGAAGGC
GCACCCCCCAGCAATCCGCGCGCCGG
GACAGAATGCCCTGCAGGAACTTCTTC
TGGAAGACCTTCTCCTCCTGCAAATAAA
ACCTCACCCATGAATGCTCACGCAAGTT
TAATTACAGACCTGAA
```

The output of the algorithm is a binary string with NxN elements, N being the dimension of the 2D barcode matrix. Using a graphical interface, the binary string can be transformed into a printable image. Each 1 value in the string will become a dot/square, while 0s will remain blank spaces. The resulting image will be squared form, with a continuous border on the left and down sides of the square and with dotted borders on the right and upper sides. The two continuous lines will be used in matrix positioning when reading the symbol.

## 4.5. Code Symbology

We agreed that two types of data will be encoded with this symbology. We will refer to these two types as **'*DNA*'** and **'*Text*'** mode and we will discuss representation for each of them. One important observation is that CWs for switching between the two modes will be needed.

## DNA mode

DNA symbols are in number of four, plus one changed symbol for RNA. That is a total of 5 symbols, easily representable on 3 bits. One extra symbol is needed for switching to text mode. For switching to

DNA mode is using 'DEL' symbol. Two CWs remain unassociated for future use:

*Table 13. DNA Symbols Encoding*

| Symbol | Coding |
|---|---|
| RFU – Reserved for Future use | 000 |
| A | 001 |
| C | 010 |
| G | 011 |
| T | 100 |
| U | 101 |
| Switch to text mode | 110 |
| RFU | 111 |

## Text mode

Text characters are encoded using ASCII standard. Although the DNA metadata only uses printable characters, we will use all the 256 symbols of the ASCII code in case we need regional languages characters and for '|' character (ASCII 179) used by some type of DNA file. The last of the 256 symbols, the one with binary representation 1111 1111, will act as "switch to DNA mode" symbol. But the 256 symbols of the ASCII code are representable on 8 bits. However, DNA metadata is reduced in length compared to the DNA data (100 text characters compared to minimum 400 DNA symbols). Therefore we will use the 3bits representation for both **DNA & Text mode**, with groups of 3 text characters being encoded over 8 * 3 bits. Total number of text will be filled with spaces until reaching a number multiple of 3.
In conclusion, we can consider a CW being 3bits in length.

A number of 400 DNA symbols will be encoded in 400 CWs on 3bits. That is 1200bits or 150Bytes. 100 text symbols will be encoded in 100 * 8 / 3 CWs, on 800 bits or 100Bytes. More important, 1bit is respended on a bar code symbol of 1,5 - 1,7 mm/character. **2000bits will therefore ocupy a square of 5/5 cm**, far less than the size of tha DNA sample file in Table 12, AND with a lot advantages in terms in **Automatic Data Capture/ Acquisitions**.

## 4.6. Error Detection

The first four CW in the matrix contain a length descriptor value, made by the number of useful information CWs. That is the number of CWs coding DNA data, added the number or correction CWs (2*k + 1, k being the correction level – please see section 4.9 – Reed Solomon corrections) and the number of error detecting CWs (one CW for length descriptor). The length descriptor can take a maximum value of $2^{12}$ = 4096 data CWs.

## 4.7. Row Structure

All CWs must be splitted into lines to form a matrix. In order to do this, one must know the number of rows and column in a matrix. Since we plan to obtain a square shaped symbol, we will use a square matrix structure. Therefore it is sufficed to calculate one single dimension of the matrix – **Ncol** – as the closest integer larger than square root of total number of CWs:

Ncol = Round(Trunc(sqrt(CW[0])))

## 4.8. Padding

Datas CWs are completed with padding CWs, non-information CWs, until being ncol*ncol in number, therefore creating all the elements of a square matrix. Padding CWs will be "000" in value.

## 4.9. Corrections: Reed-Solomon

The correction system is based on "Reed Solomon" codes as in PDF417 2D barcode – section 3 of this paper. The number of CWs to add depend of the correction level used, because of total number of CWs in a bar code (1 of which for the sum of CWs) the maximum level is limited by the number of data CWs. The number of CWs that the error correction algorithm can reconstitute is equal to the number of CWs required by the correction system. Table 8 and 9 are used for correction mechanism. Table 11 from PDF417 is replaced with Table 14:

*Table 14. Computing Correction CWs array for DNA Encoding*

Let '**s**' the correction level used, **k** = $2^{s+1}$ the number of correction CWs, '**a**' is the factors/coefficients array, '**m**' the number of data CWs, '**d**' the data CWs array and '**c**' the correction CWs array. There is a temporary-variable '**t**'.

'**c**' and '**t**' are initiated with 0. The C/C++/Java source code is:

```
for(int i = 0; < m; i++) {
        t = (d[i] + c[k - 1]) % 8;
        for(int j = k - 1; j == 0; j--)
{
        if( j == 0 ) {
        c[j] = (8 - (t * a[j]) % 8)
% 8;
        } else {
        c[j] = (c[j - 1] + 8 - (t * a[j])
% 8) % 8;
        }
        }
}

for(int j = 0; j < k; j++)
        if( c[j] != 0)
                c[j] = 8 - c[j];
```

## 4.10. Margins and Final Transformation

The final transformation of CWs consists of translating all of them into one binary string, while adding the margins that will help to position the square form. The margins are added as "1" values for continuous line and alternating "1" and "0" for dotted lines, with careful calculation of the exact position of the border values in the binary string:

- position 0 to ncol+1 for "drawing" the upper dotted line (so alternating '1' and '0' values)
- position i*ncol+2 for the left continuous line, where i goes from 1 to ncol ('1' values)
- position (i+1)*(ncol+2)-1 for the right dotted line, where i goes from 1 to ncol
- position (ncol+1)*(ncol+2) to $(ncol+2)^2$-1 for the bottom continuous line

One sample of DNA sequence generated through the presented method is shown in

Fig. 3. The symbol contains the first part of the DNA sequence coding insulin in humans:

*Table 15. DNA sample*

> insulin |homo sapiens
TACAAACATTTAGTTGTAAACACACCCTC
AGTGGACCAACTCCGCAACATAAACCAA
ACACCGCTCGCGCCGAAAAAGATATGG
GGGTTTTGG

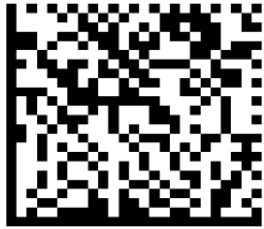

*Figure 3. DNA2DBC sample symbol*

## 5. Conclusion

There are several advantages for using 2D barcode in order to do DNA code embedded in our symbology:

- ***Improve Operational Efficiency.*** Since barcodes permit faster and more accurate recording of information, work in process can move quickly and be tracked precisely. Barcodes can respond more quickly to inquiries and changes.
- ***Save Time.*** Depending on the application, time savings can be significant. Instead of reading a code from a patient chart, type it in the computer and search it through the database, one simple symbol scan will take 2 seconds.
- ***Reduce Errors.*** Clerical errors can have a dramatic impact: wrong diagnostics, unhappy and untreated patients, and time spent to track down problems are just a few examples. Accuracy is vital in genetic diagnosis or blood bank applications. The typical error rate for human data entry is 1 error per 300 characters. Barcode scanners are much more accurate; the error rate can be as good as 1 error in

36 trillion characters depending on the type of barcode used.

- ***Save Space.*** 2D codes have the ability to encode a lot of information in a small space added to all the above advantages.
- ***Cut Costs.*** Barcodes are effective tools that can save time and reduce errors, therefore resulting in a reduction of costs.

Parts of this research have been published in the Proceedings of the 3[rd] International Conference on Security for Information Technology and Communications, SECITC 2010 Conference (printed version).